\documentclass[12pt, titlepage]{article}
\newcommand{\be}{\begin{equation}}
\newcommand{\ee}{\end{equation}}

\usepackage{epsf,rotate,psfig, latexsym}
\title{A Realistic Deterministic Quantum Theory Using 
Borelian-Normal Numbers} \author{T.N.Palmer \\ ECMWF, Shinfield Park, 
Reading\\ RG2 9AX, UK\\tim.palmer@ecmwf.int}
\begin{document}
\maketitle

\begin{abstract}

Motivated by studies of the predictability of turbulent fluids, the
elements of a deterministic quantum theory are developed, which reformulates and
extends standard quantum theory. The proposed theory is `realistic' in
the sense that in it, a general $M$-level quantum state is represented by a
single real number $0 \le r \le 1$, rather than by an element of a Hilbert space
over the complex numbers. Surprising as it may seem, this real number contains
the same probabilistic information as the element of the Hilbert space, plus
additional information from which measurement outcome is determined. A crucial
concept in achieving this is that of Borelian (number-theoretic) normality. The
essential role of complex numbers in standard quantum theory is subsumed
by the action of a set of self-similar permutation operators on the
digits and places of the base-$M$ expansion of a base-$M$ Borelian-normal $r$;
these permutation operators are shown to have complex structure and leave
invariant the normality of the underlying real number.

The  set of real numbers generated by these permutations defines not only the
Hilbert space, but also, in addition, the sample space from which quantum
measurement outcomes can be objectively determined. Dynamical real-number state
reduction is precisely described by deterministic  number-theoretic operators
that reduce the degree of normality of $r$; from the degree of normality
one can infer the standard quantum-theoretic trace rule for measurement
probability. The (probabilistic) equivalence with standard quantum theory is
demonstrated explicitly and in detail for the 2- and 3-level system.

It is suggested that many of the foundational difficulties of standard
quantum theory arise directly from the treatment of the complex Hilbert space as
axiomatic. The concepts of superposition, wave-particle duality,  state
reduction and measurement outcome, null, weak and counterfactual measurement,
the exponential speed-up of quantum computers, the many-worlds interpretation,
the classical limit of quantum theory, stochastic quantum theory, entanglement,
non-locality and the Bell inequalities, are all discussed with new insights.
These insights arise from mathematical properties of the proposed theory, rather
than from particular metaphysical assumptions. It is shown that the real-number
states $r$ of the proposed theory are precisely, using Bell's terminology, its
beables.

\end{abstract}

\section{Introduction}
\label{sec:intro}

In standard quantum theory, the quantum state is defined as an element of a
Hilbert space over the complex numbers. In this paper, it is suggested that this
theory's well-known foundational difficulties, notably the measurement problem,
arise directly from the axiomatic status of the complex Hilbert space. An
alternative theory is proposed which has no such axiom.

More specifically, the elements  of a reformulation and extension of quantum
theory are developed, each of whose states, whether associated with  a single
qubit, or the whole universe, is described by a \emph{single} real
number $0 \le r \le 1$. At first sight, such a proposal sounds preposterous.
However, $r$ has structure not usually required of a state in conventional
physical theory, and it is this structure (and deterministic but
non-arithmetic operators derived from it) that distinguishes the proposed
theory from others based, for example, on conventional differentiable
dynamics. As an illustration, a qubit or 2-level system in the
Hilbert-space state $|0\rangle+e^{i \lambda}|1\rangle$ of standard quantum
theory is described, in the proposed theory, by a single base-2 Borelian-normal
real number. By contrast, the pure Hilbert-space eigenstates $|0\rangle$ and
$|1\rangle$ are described, in the proposed theory, by the non-normal real
numbers whose binary expansions are $.000\ldots$ and $.111\ldots$ respectively.

Borelian normality is a commonplace concept in number theory, eg Hardy and
Wright(1979), though not, as mentioned, in physical theory. Roughly speaking, a
number $0 \le r \le 1$ is said to be base-2 normal if the digits 0 and 1 in the
binary expansion of $r$ appears with equal frequency, and if, additionally, the
proportions of all possible runs of binary digits of a given length are also
equal. See the footnote in section \ref{sec:complex} for more details. More
generally, for an $M$-level system, a maximally-superposed Hilbert-space state
in standard quantum theory is described by a single base-M normal real-number
in the proposed theory, whilst the pure Hilbert-space eigenstates $|0\rangle$,
$|1\rangle$, $\ldots$ $|M'\rangle$, where $M'=M-1$, are described by the $M$
non-normal real-numbers whose base-$M$ expansions are $.000\ldots$,
$.111\ldots$, $.222\ldots$, up to $.M'M'M'\ldots$.

As suggested, one of the key motivations for developing the proposed theory
was to provide a fresh approach to the measurement
problem. As discussed by Kent (2002), this is essentially the
problem of finding a  precise mathematical characterisation of the sample space
in which quantum measurement probabilities are defined.  For an
$M$-level real-number state, this sample space is defined in the proposed theory
as the countable, everywhere discontinuous set of real numbers generated by a
family of self-similar permutation operators  acting on the digits and places in
the base-$M$ expansion of some base-$M$ normal $r$. It is shown that these
operators have a well-defined complex strucuture (equivalent to $M$th roots of
unity), and it is this complex structure which subsumes the essential role of
complex numbers in defining the Hilbert-space state in standard quantum theory.
These permutation operators, discussed in detail in sections \ref{sec:complex}
and \ref{sec:3l} for $M=2$ and $M=3$ respectively, define the
heart of the paper.

In addition to these self-similar permutation operations (which leave
invariant the Borelian normality of the associated real number), a class of
number-theoretic operators are defined which reduce the
degree of normality of $r$ (see sections \ref{sec:nonnormal} and \ref{sec:3l}).
For example, if $r=.02110211\ldots$ is the base-3 expansion of a base-3 normal
number, then $r_{\perp 2}=.011011\ldots$ (ie deleting all occurrences of the
digit `2'), defines the binary expansion of a base-2 normal number. These
reduction operators are equivalent to the Hilbert-space projection operators of
standard quantum theory. However, these number-theoretic reduction operators
allow quantum-state reduction to be defined as a process governed by a precise
deterministic mathematical procedure (`$R$'). In this way, the proposed theory
extends standard quantum theory.  For example, with $|0\rangle+e^{i
\lambda}|1\rangle$ described by the base-2 normal real $r$, then, under $R$, $r
\mapsto r_{\perp1 }= .000\ldots$ if $r<1/2$, and $r \mapsto r_{\perp 0}=
.111\ldots$ if $r \ge 1/2$. The numbers $.000\ldots$ and $.111\ldots$ are fixed
points of $R$. Under $R$, the state space of a real-number qubit state can be
described as comprising two arbitrarily-intertwined basins of attraction of
point attractors $\mathcal{A}_0$ and $\mathcal{A}_1$, where $r=.000\ldots$ and
$r=.111\ldots$ respectively.

In sections \ref{sec:dynamics} and \ref{sec:3l},  the probabilistic equivalence
between the Hilbert-space description of standard quantum theory and the
real-number state of the proposed theory,  is discussed in detail for  2- and
3-level systems . For the qubit, the real number equivalent of the Bloch sphere
is constructed explicitly. It is shown that quantum measurement probabilities in
the proposed theory satisfy the trace rule. The equivalent of some explicit
unitary transformations in standard theory are given in section
\ref{sec:unitary}. The (instrinsically irreversible) properties of such
number-theoretic reduction operators are consistent with properties of quantum
gravity, as speculated by Penrose (1989, 1994, 1998). The notions of
quantum-state superposition, polarisation and interference are discussed in
section \ref{sec:unitary}, from the perspective of the proposed theory. As
discussed in section \ref{sec:measurement}, the proposed theory provides a
simple and novel relationship between the notions of reduction and measurement
outcome, without recourse to some arbitrary classical/quantum split; in the
proposed theory, an observer (or indeed the whole cosmos) is merely an $M$-level
system where $M \gg 1$.  A brief description of the (non-singular) classical
limit of the proposed theory is given.

The formalism developed in this paper was motivated by studies of the
predictability of the Navier-Stokes equations for turbulent fluids in
meteorology (Palmer, 2000). Exploiting this, the concept of a `Navier-Stokes'
computer is introduced in section \ref{sec:computing} to show how the
proposed theory provides an alternative single-world view to the many-worlds
explanation of  the exponential speed-up of certain quantum computational
problems over their digital counterparts.

One of the key features of the sample space comprising the reals constructed by
the family of self-similar permutations, is that it is everywhere discontinuous
and, for a qubit, is only definable on a countable set of meridians of the
Bloch-sphere equivalent. As discussed in section \ref{sec:bell}, this has
important physical consequences for the proposed theory, most importantly that
counterfactual measurements, whose outcomes are not elements of reality by
definition, cannot be associated with well-defined real-number states. In this
sense, the proposed theory is fundamentally different from any classical
deterministic theory. With such counterfactual indefiniteness, it is shown that
the theory can violate the Bell inequalities, and yet be local, at least in the
EPR (Einstein et al, 1935) sense of the word.

In summary, for any mathematically well-defined real-number state of the
proposed theory, its transform under reduction (and therefore measurement) is
determined precisely. Conversely, measurements which by definition cannot be
elements of reality, are not associated with real-number states. Hence
real-number states and elements of reality are in one-to-one correspondence in
the proposed theory. It is for this reason that the real-number states $r$ of
the proposed theory that include both microscopic particles and macroscopic
observers, are precisely its beables.\footnote{To quote Bell(1993): ``The
beables of the theory are those elements which might correspond to elements of
reality, to things which exist. Their existence does not depend on
`observation'. Indeed observation and observers must be made out of beables.''}

 \section{Complex structure from self-similar permutations}
\label{sec:complex}

Let
\be
\mathcal{S}_r=\{a_1,a_2,a_3, \ldots\},
\ee
$a_i \in \{0,1\}$, denote a bit string whose elements  define the binary
expansion \be
r=.a_1a_2a_3\ldots
\ee
of some real number $0 \le r \le 1$. Let $\phi(0)=1,  \phi(1)=0$, and
\be
-\mathcal{S}_r = \{\phi(a_1), \phi(a_2), \phi(a_3), \ldots\},
\ee
so that $-(-\mathcal{S}_r)=\mathcal{S}_r$.
Defining
\begin{eqnarray}
i(\mathcal{S}_r) &=&\{\phi(a_2), a_1, \phi(a_4), a_3, \phi(a_6), a_5, \phi(a_8),
 a_7, \ldots \} \nonumber \\
i^{1/2}(\mathcal{S}_r) &=& \{\phi(a_4), a_3, a_1, a_2, \phi(a_8), a_7, a_5,a_6,
 \ldots\}
\end{eqnarray}
then it is easily shown that  $i*i(\mathcal{S}_r)=-\mathcal{S}_r$, and
 $i^{1/2}*i^{1/2}(\mathcal{S}_r)=i(\mathcal{S}_r)$.
The operators $i$ and $i^{1/2}$ induce the real-number transformations
\begin{eqnarray}
r \mapsto \tilde{i}(r)&=&.\phi(a_2) a_1 \phi(a_4) a_3 \phi(a_6) a_5 \phi(a_8)
a_7 \ldots \nonumber \\
r \mapsto \tilde{i}^{1/2}(r)&=& .\phi(a_4) a_3 a_1 a_2 \phi(a_8) a_7 a_5a_6
\ldots \end{eqnarray}

This construction is easily generalised to a family of
permutation operators $i^{1/2^n}$ (integer $n \ge 0$) such that
\be
\label{eq:ss}
i^{1/2^n}*i^{1/2^n}(\mathcal{S}_r)=i^{1/2^{n-1}}(\mathcal{S}_r)
\ee
To do this, it will be convenient to write
\be
\label{eq:tuplet}
\mathcal{S}_r=\{\beta^{(n)}_1,\beta^{(n)}_2, \beta^{(n)}_3, \ldots \}
\ee
where $\beta^{(n)}_j$ is the $j$th contiguous $2^n$-tuplet of elements of
 $\mathcal{S}_r$, ie
\be
\beta^{(n)}_j=\{a_{(j-1)2^n+1}, a_{(j-1)2^n+2}, \ldots, a_{j 2^n}\}.
\ee
Consider an operator $\chi^{(n)}$ which permutes the digits and the places of
the elements of each $\beta^{(n)}_j$ as follows: operate on the last
element of $\beta^{(n)}_j$ with $\phi$,  then swap the last two contiguous
elements of $\beta^{(n)}_j$, then swap the  last two contiguous pairs of
elements of $\beta^{(n)}_j$, then swap the last  two contiguous quadruplets of
elements and so on, finally swapping the  two contiguous $2^{n-1}$-tuplets of
elements of $\beta^{(n)}_j$.

$\chi^{(n)}$ can be expressed more succinctly as the $2^n \times 2^n$ matrix
defined by the iterative self-similar  block form
\be \label{eq:block}
\chi^{(n)}=\left(
 \begin{array}{cc}
 0&I\\
\chi^{(n-1)}&0
\end{array}
\right)
\ee
where $I$ is the $2^{n-1} \times 2^{n-1}$ identity matrix and
$\chi^{(0)}=\phi$. The matrix acts on each $\beta_j^{(n)}$, considered as a row
vector.  For example, from equation \ref{eq:block}, the effect of $\chi^{(3)}$
on $\beta_1^{(3)}$ is represented by
\begin{eqnarray}
\left(
\begin{array}{cccccccc}
a_1&a_2&a_3&a_4&a_5&a_6&a_7&a_8
\end{array}
\right)
\left(
\begin{array}{cccccccc}
0&0&0&0&1&0&0&0\\
0&0&0&0&0&1&0&0\\
0&0&0&0&0&0&1&0\\
0&0&0&0&0&0&0&1\\
0&0&1&0&0&0&0&0\\
0&0&0&1&0&0&0&0\\
0&1&0&0&0&0&0&0\\
\phi&0&0&0&0&0&0&0
\end{array}
\right)\nonumber\\
=\left(
\begin{array}{cccccccc}
\phi(a_8)&a_7&a_5&a_6&a_1&a_2&a_3&a_4
\end{array}
\right)
\end{eqnarray}
Based on these matrix operations, define
\be
\label{eq:in}
i^{1/2^{n-1}}(\mathcal{S}_r)=\{\beta'^{(n)}_1,\beta'^{(n)}_2,\beta'^{(n)}_3,
\ldots \} \ee where the permuted $2^n$-tuple $\beta'^{(n)}_j$ is given by the
elements of the row vector $\beta^{(n)}_j \chi^{(n)}$. Equation \ref{eq:ss}
immediately follows from the identity
\be
 \left(
 \begin{array}{cc}
 0&I\\
\chi^{(n-1)}&0
\end{array}
\right) \times
\left(
 \begin{array}{cc}
 0&I\\
\chi^{(n-1)}&0
\end{array}
\right)=
\left(
 \begin{array}{cc}
 \chi^{(n-1)}&0\\
 0&\chi^{(n-1)}
\end{array}
\right)
\ee
For any dyadic rational $q=m/2^n$, integer $m$ $n$, one can therefore define
\be
i^q(\mathcal{S}_r)=\underbrace{i^{1/2^n}*i^{1/2^n} \cdots
 i^{1/2^n}}_m(\mathcal{S}_r).
\ee
which in turn induces the real-number transformation $r \mapsto
\tilde{i}^q(r)$.

The numbers $r$ on which $\tilde{i}^q$ operate are now considered. We first
note that $\tilde{i}^q$ preserves Borelian normality. It is well known that
almost all real numbers are normal (eg Hardy and Wright, 1979)\footnote{The
concept of normal numbers was introduced by Emile Borel. Suppose the  digit $b
\in \{0,1,2\ldots M-1\}$ occurs $n_b$ times in the first $n$ places of the the
base-$M$  expansion of some real $r$. If, for all b, $n_b/n \rightarrow 1/M$ as
$n \rightarrow \infty$, then $r$ is simply normal in base  $M$. If $r$ is simply
normal in all of base $M, M^2,M^3 \ldots$, then $r$ is said to be  normal in
base $M$. Normality is a generic property of the reals, yet it is still unknown
whether numbers such as $\pi$ and $e$ are normal in any base. In the discussion
below this notion of normality is generalised slightly: $r$ will be said to be
normal in base $M$, if it satisfies the conditions above, irrespective of the
actual symbols used to represent the $M$ digits $b$. Hence, for example,
$.000\ldots$, $.111\ldots$, $.222\ldots$ and so on, will all be described as
base-1 normal.}. If $r$ is base-2 normal in $\{0,1\}$, then there is an equal
probability of finding a `0' or a `1' at a given place in the binary expansion
of $r$. Hence, the application of $\phi$ does not change the probability of
finding a `0' or `1' at the given place. Similarly none of the swap permutations
in the definition of $\chi^{(n)}$ alter the fraction of `0's and `1's in each
$\beta^{(n)}_j$.  Hence, if $r$ is a base-2 normal real, so is
$\tilde{i}^q(r)$.

 Let $r_0$ denote some arbitrary base-2 normal real; a possible choice  is the
binary version $r_C=0.011011100101\ldots$ of Champernowne's famous normal
number defined by concatenating the integers (eg Hardy and Wright, 1979). With
angular coordinate $\lambda = 2\pi q$, and $q$ a dyadic rational, we define  \be
\label{eq:fn}
r(\lambda)=\tilde{i}^{4q}(r_0)
\ee
so that $r(0)=r_0$ and each $r(\lambda)$ is base-2 normal.
We now \emph{define} the permutation operator $e_{p}^{2 i \pi q}$ acting
on arbitrary bit strings $\mathcal{S}_r$ by
 \be
 \label{eq:e}
 e_{p}^{2 i\pi q} (\mathcal{S}_r)= i^{4q} (\mathcal{S}_r)
 \ee
Hence equation \ref{eq:fn} can be written in the real-number `phase' form \be
\label{eq:expo} r(\lambda)=e_{\tilde{p}}^{i \lambda}(r_0) \ee where
$e_{\tilde{p}}^{i \lambda}$ is the real-number operator induced by
$e_{p}^i$. Note that, where used in this paper, the expression $e^{i
x}$ continues to denote the complex exponential of conventional
mathematics.

Note that, by normality, every conceivable bit string $\mathcal{B}_N$ of length
$N$  can be expected to occur once in the first $2^N$ places in the binary
expansion of any $r(2\pi q)$. As discussed below, this is the basis of
the equivalence between the notion of superposition in the Hilbert-space
formulation of standard quantum theory, and the real-number formulation of the
proposed theory. Similarly, for any base-2 normal $r_0$, there exists  a
$\lambda$ which is a dyadic rational multiple of $\pi$, such that the binary
expansion of $r(\lambda)$ begins with $\mathcal{B}_N$. Hence, the transformation
$r_0 \mapsto r'_0$ between base-2 normals, can be effected by some $\lambda
\mapsto \lambda'$ with fixed $r_0$. The  independence of measurement statistics
to the particular choice of $r_0$ is equivalent in standard quantum theory to
the independence of measurement statistics to the choice of global phase factor
in the standard qubit  representation (cf equation \ref{eq:state}).

For the discussion on the origin of exponential speed up in quantum computing in
section \ref{sec:computing}, on counterfactual indefiniteness and quantum
nonlocality in section \ref{sec:bell},  and on the concept of weak measurement
and its relation to stochastic quantum theory in section \ref{sec:stochastic},
it is essential to note that  $r(\lambda)$ does not vary continuously or
monotonically with $\lambda$; indeed $r(\lambda)$ is not defined when
$\lambda$ is not a dyadic rational multiple of $\pi$. To see this, let $\Delta q
= 1/2^n$ and note that, by equation \ref{eq:fn}, $\mathcal{S}_{r(2 \pi \Delta
q)}=i^{4 \Delta q}(\mathcal{S}_{r(0)})$. By definition, the first element
of $\mathcal{S}_{r(2 \pi \Delta q)}$ is equal to the $\phi$ permutation of the
$2^{n-1}$th element of $\mathcal{S}_{r(0)}$. Hence the smaller is $\Delta q$,
the  further back in the string $\mathcal{S}_{r(0)}$ is drawn the first element
of  $\mathcal{S}_{r(2 \pi \Delta q)}$. As a result, for
arbitrarily small $\Delta \lambda$, the sequence of numbers $\{r_1, r_2,
\ldots\}$  where $r_j=r(\lambda + j\Delta \lambda)$ has a lag-1 correlation of
zero, that is, the sequence appears as if it were stochastic white noise. As a
result, it is claimed in section \ref{sec:stochastic} that the proposed theory
can emulate some of the features of stochastic quantum theory, without actually
being stochastic itself, nor being chaotic in the conventional dynamical-system
sense.

All normal numbers are irrational. However, no aspect of the theory presented
 below requires us to consider strictly irrational numbers. Indeed, as discussed
 below, there may be physical constraints which limit
 the bit string length of the binary expansion of $r_0$.  Let $\hat{r}_0$ denote
 a  rational approximation to $r_0$ in the sense that the digits in the first
$2^N$ places in the binary expansions of $\hat{r}_0$ and $r_0$ agree. Then, with
$\hat{r}(0)=\hat{r}_0$, dyadic rational numbers $\hat{r}(2 \pi j/2^N)$, where
$j=1,2,3 \ldots 2^N$ can be defined at a finite set of $2^N$ points on the
circle, with a fundamental longitudinal spacing $2\pi/2^{N}$. Each $\hat{r}(2
\pi j/2^N)$ is a  rational approximation of the corresponding normal number
$r(2\pi j/2^N)$ in the sense that the bits in the first $2^N$ places of their
binary expansions agree. In the analysis that follows, it can be assumed that we
are dealing with  such rational approximations, rather than  formally irrational
numbers. As such the proposed theory is essentially finite,  and in particular
does not utilise any of the paradoxical properties of non-measurable infinite
sets (eg Pitowsky, 1983).

\section{A Number-Theoretic Reduction Procedure}
 \label{sec:nonnormal}
In this section, the second essential non-arithmetic operation
for constructing real-number states is proposed: number-theoretic
reduction. As a simple example of this concept, the
`reduced' number $r_{\perp 0}(\lambda)=.111\ldots$ arises by deleting  all
places where the digit `0' occurs, in the binary expansion of $r(\lambda)$.
Similarly $r_{\perp 1}(\lambda)=.000\ldots$ arises by deleting all places where
the digit `1' occurs. More generally, a number-theoretic procedure is
defined below which generates a one-parameter family of numbers starting with
the base-2 normal $r(\lambda)$ and finishing with either of the base-1 normals
$.000\ldots$ or $.111\ldots$. As discussed below, since the degree of normality
of these numbers decreases monotonically with this parameter, the procedure is
referred to as `number-theoretic reduction'.

To define this procedure explicitly, and to relate with the Bloch sphere of
standard quantum theory, let $E$ denote the equator, and $p_N$ and $p_S$
the corresponding poles, of a two sphere $S^2$. Let $\theta$ denote co-latitude,
and $\Lambda$ denote the set of meridians whose longitude $\lambda$ is a
dyadic rational multiple of $2\pi$. Start with the bit string
$\mathcal{S}=\{a_1a_2a_3\ldots\}$ defined from the binary expansion of
$r(\lambda)$, and then delete some of the digits $a_j$ based
on the following rules. If $a_j=1$, then delete $a_j$ from $\mathcal{S}$ iff
\be \label{eq:pp1} .a_ja_{j+1}a_{j+2}\ldots < \cos^2
\frac{\theta}{2}
\ee
Conversely, if $a_j=0$ then delete $a_j$ iff
\be
\label{eq:pp2}
.a_ja_{j+1}a_{j+2}\ldots \ge \cos^2
\frac{\theta}{2}
\ee
The real-number function $r(\theta, \lambda)$ is defined from its binary expansion \be
r(\theta, \lambda)=.a'_1a'_2a'_3\ldots
\ee
where the $a'_j$s correspond to the digits which have not been deleted from
$\mathcal{S}$.

For example, at $\theta=0$ all occurrences of $a_j=1$ and no occurrences of
$a_j=0$ are deleted. At $\theta=\pi/2$, none of the $a_j$ are deleted.
At $\theta=\pi$, all occurrences of $a_j=0$ are deleted.  Hence
\begin{eqnarray} \label{eq:fn2} r(0,\lambda)&=&.000\ldots\nonumber\\
r(\pi/2,\lambda)&=&r(\lambda)\nonumber\\ r(\pi,\lambda)&=&.111\ldots
\end{eqnarray}
More generally, based on an ensemble of numbers $r(\theta, \lambda)$
obtained by sampling $\lambda$ (from $\Lambda$),  $P[r(\theta, \lambda)
< 1/2]=  \cos^2 \theta/2$, where $P$ is a probability measure.

To see this, note that if $r(\theta, \lambda) < 1/2$, then the first digit in
the binary expansion of $r(\theta, \lambda)$ must be a zero. According to the
reduction procedure above, $a'_1=0$ if  $r(\lambda) < \cos^2
\theta/2$. Hence
\be P[r(\theta, \lambda) < 1/2]=P[r(\lambda) < \cos^2 \theta/2]
\ee The required result now follows from the fact that for
any base-2 normal number $0 < r < 1$ and any $\theta$, then $P[r <
\cos^2\theta/2]=\cos^2\theta/2$. (To show this, let the binary expansion of
$\cos^2 \theta/2$ be $.c_1c_2c_3\ldots$. Suppose $c_j=0$ except where
$j=j_1,\ j_2,\ j_3 \ldots$. Hence, if $r < \cos^2\theta/2$, then $a_1,\ a_2,
\ldots a_{j_1-1}$ must all be equal to 0. If, in addition, $a_{j_1}$ is 0, then
certainly $r < \cos^2\theta/2$. Since $r$ is base-2 normal, the probability of
`0's in all of the first $j_1$th positions in the binary expansion of $r$ is
equal to $1/2^{j_1}$ or, in binary notation, $.000\ldots1$ where the `1' occurs
in the $j_1$th place. By definition, this probability is equal to $.c_1c_2\ldots
c_{j_1}$.  Now in addition, $r<\cos^2\theta/2$ if $a_{j_1}=1$, but all
other $a_i$ up to and including the $j_2$th element are equal to 0.  The
probability of such additional occurrences is (in binary) $.000\ldots1$ where
now the `1' occurs in the $j_2$th place. Continuing this agument, the total
probability that $r<\cos^2\theta/2$ is equal to  \be
\underbrace{.000\ldots1}_{j_1}+\underbrace{.000\ldots1}_{j_2}+
\underbrace{.000\ldots1}_ {j_3}+ \dots  \ee  where the underbrace value gives
the length of the binary expansion. By definition this sums to
$.c_1c_2c_3 \ldots =\cos^2\theta/2$ QED.

The quantity $\cos^2 \theta/2$ which on the one had defines the probability
that $r(\theta, \lambda) < 1/2$, can also be viewed as defining the `degree of
normality' of $r(\theta, \lambda)$, (the relative fraction of `0's and `1's
in the binary expansion of $r$), with maximal base-2 normality at
$\theta=\pi/2$, minimal base-2 normality at $\theta=0,\ \pi$. As discussed
below, this links in a fundamental way in the proposed theory, the probability
of measurement outcome with the degree of normality of the underlying
real-number quantum state.

Finally, note that if the value of $r$ at any point $p$ is $r(p)$,
then the value of $r$ at the point $p'$ antipodal to $p$ is $r(p')=1-r(p)$. See
Figure  1.

\begin{figure}
\epsfxsize=10cm
\centerline{\epsffile{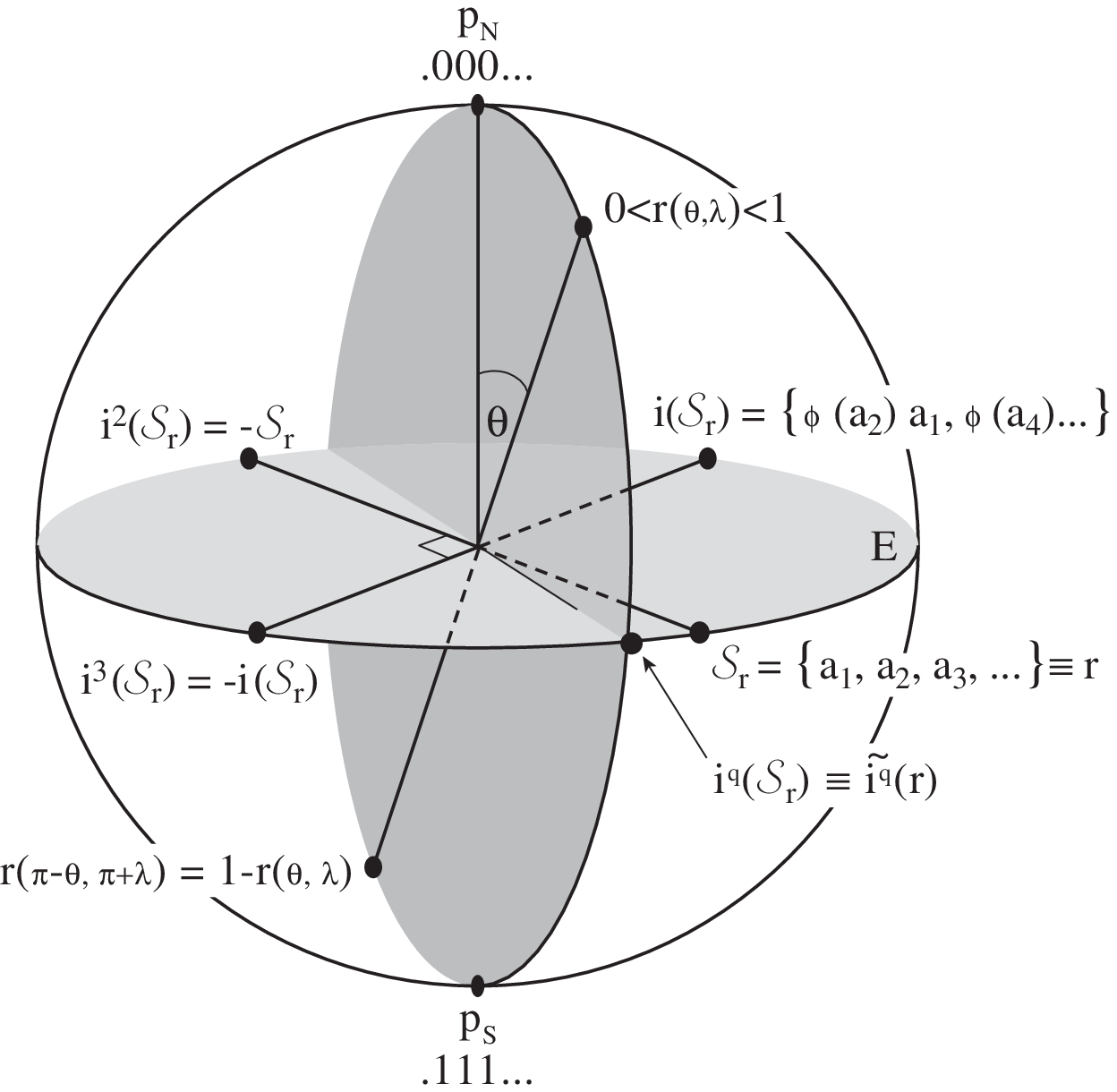}}
\caption{\textsf{In the proposed theory, a state of a qubit is a
real number $0 \le r \le 1$ defined on a countable subset $\Lambda$ of meridians
on the two-sphere $S^2$. On the equator, $r$ is base-2 normal, defined from the
family of self-similar permutation operators $i^q$ (with $q$ a dyadic rational)
acting on some bit string $\mathcal{S}_{r}$. The further away from the equator,
the less $r$ is base-2 normal. The north and south poles, where $r=.000\ldots$
and $r=.111\ldots$ respectively, correspond to minimally-normal fixed points of
a deterministic reduction operator $R$. Under $R$, an arbitrary real-number
state $r$ is attracted to one or the other of these fixed points, depending on
whether $r < 1/2$ or $r \ge 1/2$. In turn this determines, for example, whether
a measurement outcome will be `spin up' or `spin down'. The orientation of the
poles is determined by the qubit's coupling to some $M$-level system, $M \gg
1$, see section \ref{sec:measurement}}} \label{fig:Fig1} \end{figure}

\section{Relation to 2-level Quantum Theory}
\label{sec:dynamics}

The central thesis of this paper is that $r$, as defined in the previous two
sections, is to be viewed as a real-number state in a deterministic theory which
subsumes standard quantum theory. The symbol `$\lhd$' is used to represent
this thesis inside mathematical equations. Hence, for the 2-level quantum
state, the subsumption \be \label{eq:state} |\psi\rangle = e^{i \gamma}(\cos
\frac{\theta}{2}\ |0\rangle + e^{i \lambda} \sin \frac{\theta}{2}\  |1\rangle)
\lhd r(\theta, \lambda) \ee is proposed between the standard quantum-theoretic
Bloch-sphere representation of the Hilbert-space qubit state, and the
real-number state $0\le r(\theta, \lambda) \le 1$.

However, if one is only concerned with the statistics of measurement outcome,
as is the case in practice in quantum experimentation, then it is claimed that
the proposed theory is equivalent to standard quantum theory. In this sense, the
symbol $\lhd$ also denotes an equivalence between the
Hilbert-space state and the real-number state at the level of probabilities.
Since $|0\rangle$ and $|1\rangle$ denote Hilbert-space eigenstates of some given
Hermitian operator $\mathcal{O}$ (eg associated with spin in some prescribed
direction), this probabilistic equivalence is specific to the choice of operator
(or `observable'). The number $r$ is referred to as the `real-number
$\mathcal{O}$-state', or , hereafter, `real-number state' of the two level
system.

The issue of what determines the orientation of the polar axis
in a theory which purports to construct observables from beables (the basis
problem) is addressed in section \ref{sec:measurement}. For the purposes of this
section, the orientation of the axis are assumed given.

\subsection{Global Phase Factor}

As is well known, in standard quantum theory, the statistics of measurement
outcome do not depend on the choice of global phase factor $\gamma$ in equation
\ref{eq:state}. In the proposed theory, the choice of global phase factor
$\gamma$ is equivalent to the choice of base-2 normal $r_0$ associated with the
zero of longitude. For example, if we write
\be
|0\rangle+|1\rangle \lhd r(\pi/2,0)=r_0
\ee
then
\be
e^{i \gamma}(|0\rangle+|1\rangle) \lhd r'(\pi/2,0)=\tilde{i}^{2
\gamma/\pi}(r_0)
\ee
As discussed above, for any $r_0$, there exists a $\gamma$
commensurate with $\pi$ such that $\tilde{i}^{2 \gamma /\pi}$ starts with
any given finite bit string. Hence, as discussed below, the independence of
the statistics of measurement outcome to the choice of $r_0$ is equivalent to
the fact that the statistics of $r(\theta, \lambda)$ are independent of
$\lambda$.

\subsection{Number-Theoretic Reduction and Hilbert-Space Projection}

Reduced numbers are an important component of the proposed theory. For example,
if
\be
\frac{1}{\sqrt{2}}(|0\rangle+|1\rangle) \lhd r(\pi/2,0)
\ee
is a base-2 normal real, then
\begin{eqnarray}
|0\rangle \lhd r_{\perp 1} (\pi/2,0)&=&.000\ldots\nonumber \\
|1\rangle \lhd r_{\perp 0} (\pi/2,0)&=&.111\ldots
\end{eqnarray}
are base-1 normal reals, and more generally,
\be
\cos \frac{\theta}{2}\ |0\rangle + \sin
\frac{\theta}{2}\  |1\rangle \lhd r(\theta, \lambda)
\ee
is a partially-reduced real number, (ie neither base-2-normal nor base-1-
normal). As discussed in section \ref{sec:nonnormal}, the degree of normality of
$r$ is defined to be equal to $\cos^2 \theta/2$. (The generic importance of
reduced numbers will become more apparent when we consider 3-level and $M$-level
systems below.) It can be noted that the concept of number-theoretic
reduction in the proposed theory is equivalent to the concept of
Hilbert-space state projection in conventional quantum theory (eg $r \mapsto
.000\ldots$ is equivalent to $|0\rangle+|1\rangle \mapsto |0\rangle$). However,
unlike conventional quantum theory, reduction is a deterministic process,
internal to the dynamics of the proposed theory. This dynamical process is
described in the next subsection.

\subsection{Dynamical Real-Number State Reduction}
\label{sec:reduction}

One of the principal features of the proposed theory is that quantum state
reduction is an objective process defined by a deterministic mathematical
procedure.  Moreover the theory provides a precise formulation of the sample
space on which quantum reduction probabilities are defined.  In this section we
describe this reduction criterion on a qubit; its generalisation to a 3-level
system is described in section \ref{sec:3l}. In section \ref{sec:measurement} we
show how real-number state reduction can be simply  and deterministically linked
to the notion of measurement outcome, without recourse to any artificial
`classical/quantum' split. Since the state is always a definite real number, the
problem is never faced of `when', during quantum state reduction, the notion of
a complex superposition of outcomes is no longer relevant: in the proposed
theory, it is never relevant. As a result, as discussed in section
\ref{sec:measurement}, the proposed model has a simple non-singular classical
limit.

The process of real-number state reduction is based on the number-theoretic
notion of reduction discussed in section \ref{sec:nonnormal} above. For the
qubit, we first consider the reduction operation $R_j:[0,1] \rightarrow [0,1]$,
$j \in \{0,1\}$. Specifically, if \be \label{eq:reduce} r=.a_1a_2a_3\ldots
 \ee
is the binary expansion of some qubit real-number state $r$, then
\begin{eqnarray}
R_j: r \mapsto .jjj\ldots \ \mathrm{if}\ a_1=j \nonumber \\
R_j = \mathrm{Identity}\ \mathrm{if}\ a_1 \ne j
\end{eqnarray}
Hence under $R_1$, the real-number state evolves to $p_S$ if $r \ge 1/2$ and
remains unchanged if $r < 1/2$. Under $R_0$, the real-number state evolves to
$p_N$ if $r < 1/2$ and remains unchanged if $r \ge 1/2$. The real-number states
$.000\ldots$ and $.111\ldots$ are fixed points of $R_0$ and $R_1$.  We can form
the compound reduction operator $R=R_0R_1$. Under $R$ the real-number state
evolves to $p_S$ if $r \ge 1/2$, and to $p_N$ if $r < 1/2$. For example, $R_1$
could be used to ask the question: `was a particle detected', whilst $R$ could
be used to ask the question: `was the particle spin up or spin down'? A null
measurement (`particle not detected') leaves the real-number state unchanged,
but clearly does not imply that the real-number state is equal to zero.

It is easy to suggest a differential equation to describe the time
evolution of the real-number state under $R$. For example, let
\begin{eqnarray}
\label{eq:ode}
\dot{\theta}&=&\alpha\ (r-\frac{1}{2}) \sin \theta \nonumber \\
\dot{\lambda}&=&0
\end{eqnarray}
where $r=r(\theta, \lambda)$, and  $\alpha$ is a parameter discussed in section
\ref{sec:measurement} which sets the fundamental timescale for reduction. During
the reduction process (as described by equation \ref{eq:ode}), the degree of
normality of the real-number state $r$ reduces continuously and irreversibly.
In this way, $p_N$ and $p_S$ can be considered attractors $\mathcal{A}_0$ and
$\mathcal{A}_1$ respectively, of an irreversible deterministic dynamical system.
Effectively, $R$ defines a colouring of $\Lambda$, eg $p$ could be coloured red
iff $r(p)\mapsto .000\ldots$ under $R$, blue iff $r(p) \mapsto .111\ldots$
under $R$. Given the properties of $r$ then if $p$ is coloured red, the
antipodal point $p'$ is coloured blue (and vice versa). If $p \notin \Lambda$
then $p$ is not coloured.  Hence, red points  lie in the basin of attraction of
$\mathcal{A}_0$;  blue points lie in the basin of attraction of $\mathcal{A}_1$.
Because of the discontinuous and non-monotonic nature of $r$, then there are
blue points in any $\lambda$ interval of any red point (and vice versa).
Effectively equation \ref{eq:ode} describes a deterministic intertwined-basin
dynamical system (Ott et al, 1993; Palmer, 1995; Duane, 2001; Nicolis et al,
2001). However, because of the non-differentiable nature of $r$, \ref{eq:ode} is
qualitatively unlike any intertwined-basin system based on conventional chaotic
dynamics. This qualitative difference with conventional deterministic systems
will be particularly relevant when the reason why the proposed theory can
violate Bell's theorem is discussed (section \ref{sec:bell}).

As described in section \ref{sec:measurement} below, the general real-number
state of an $M$-level system is described by a real $0 \le r \le 1$ which is at
most base-$M$ Borelian normal.  In general, it is postulated that the effect of
the reduction operator on the real-number state of an isolated system, is to
reduce monotonically the state's degree of Borelian normality. Hence if $M$ is
large, and the system at time $t$ is well approximated by a base
$M(t)$-normal real, then, in the proposed theory, then $M(t_1)<M(t_0) \le M$,
where $t_1>t_0$. This number-theoretic reduction process cannot derive from
existing theories of physics, none of which attributes any significance to the
degree of normality of the state, and none of which is irreversible in the sense
described. We must therefore seek some fundamental physical process which may be
consistent with such a number-theoretic description. Quantum gravitation (whose
dynamical formulation is still open to debate) appears the only candidate. The
arguments for gravity being crucial to quantum-state reduction have been given
eloquently by Penrose (1989, 1994, 1998) and will not be repeated here. However,
note in particular that Penrose provides support for the notion that, unlike the
other forces of physics, the gravitational force is time asymmetric (see also
Prigogine and Elskens, 1987; 't Hooft, 1999).

\section{Wave-Particle Duality}  \label{sec:unitary}

\subsection{Schr\"{o}dinger Dyanamics}

Consider a source emitting a single qubit every time $\Delta t$. Let the real-number state
at time $t_n$ be given by $r(t_n)$ with  corresponding bit string
$\mathcal{S}(t_n)$. Suppose \be
\mathcal{S}(t_n)=i^{4\nu n \Delta t}(\mathcal{S}(t_0))
\ee
That is, the real-number state at time $t_n$ is obtained from the real-number state at time $t_{n-1}$
by a (phase) rotation through $2 \pi \nu \Delta t$ about the polar axis.
Using the permutation operator $e_p^{i \lambda}$ defined in equations \ref{eq:e}
and \ref{eq:expo}, then with $\omega=2 \pi \nu$, we can write
\be
\label{eq:sch}
r(t_n)=e_{\tilde{p}}^{i n \omega \Delta t}(r(t_0))
\ee
Compare this with wavefunction evolution in the conventional form
\be
\label{eq:sch2}
|\psi(t_n)\rangle=e^{i n \omega \Delta t}|\psi(t_0)\rangle
\ee
Hence, in the proposed theory, $\mathcal{S}(t_n)$ describes a coherent
monochromatic wave  source, with frequency $\omega$ and intensity $1/\Delta t$.
With low enough intensity, individual qubits (`particles') can be measured. With
the basic quantum premise that $\mathcal{E}=\hbar \omega$, then \ref{eq:sch} is
equivalent to a first integral of the Schr\"{o}dinger equation for a
free-particle with energy $\mathcal{E}$.

\subsection{Polarisation and Uncertainty}
Imagine an ensemble of real-number states associated with points in some small neighbourhood
of $(\theta, \lambda)$, so that the real-number state of the $j$th qubit in this
prepared ensemble is
 \be
\label{eq:polarisation}
r(\theta_j, \lambda_j)=r(\theta +\delta\theta_j, \lambda+\delta \lambda_j) \ee
Now by section \ref{sec:nonnormal}, the numerical values of the real-number
states in equation \ref{eq:polarisation} are not sensitive to small variations
$\delta\theta_j$, but by section \ref{sec:complex}, they are sensitive to small
variations $\delta\lambda_j$.  As discussed in section \ref{sec:nonnormal}, the
probability that one of the qubits in this ensemble reduces to $p_N$ is equal to
the probability that this real-number state is $< 1/2$, which equals $\cos^2 \theta/2$,
consistent with quantum polarisation statistics for prepared states. When
$\theta = \pi/2$, then, from this probabilistic perspective, there  is complete
uncertainty as to whether a randomly-chosen real-number state will reduce to
$p_{N}$ or $p_{S}$, in no matter how small a neighbourhood of real-number
state space the ensemble of states is initially prepared.

\subsection{Unitary Transforms and Superposition}

For the two level system, it is well known in standard quantum theory that
unitary transformations $U:|\psi\rangle \rangle \mapsto  |\psi'\rangle$  are
associated with the action of arbitrary isometries of the Bloch sphere.  In the
proposed theory, such isometries define the equivalent  unitary-like
transformations $\tilde{U}:r(\theta, \lambda) \mapsto r'(\theta, \lambda)$.

As an example, the unitary transform
\begin{eqnarray}
|0> &\mapsto& \frac{|1\rangle+|0\rangle}{\sqrt{2}} \nonumber \\
|1> &\mapsto& \frac{|1\rangle-|0\rangle}{\sqrt{2}}
\end{eqnarray}
in standard quantum theory, corresponds, in the proposed theory,  to the
 $\tilde{U}$ transform
\begin{eqnarray}
\label{eq:had}
.000\ldots  &\mapsto& r_0 \nonumber \\
.111\ldots &\mapsto& \tilde{i}^2(r_0)=1-r_0
\end{eqnarray}
Note that in the proposed theory, each of the states $r_0$ and $1-r_0$ is a
definite real number, not some complex-number weighted coexistence of
alternative eigenstates. The equivalence with the notion of superposition
arises from the fact that in equation \ref{eq:had},  the binary expansions of
the base-2 normal reals $r_0$ and $1-r_0$, each contain equal numbers of `0's
and `1's.

\subsection{Interference}

Imagine a beam of photons impinging on a half-silvered mirror. Standard
quantum theory views as fundamentally non-local, the Hilbert-space state
$|0\rangle +i|1\rangle$ of the photon after passing  through the half-silvered
mirror, where $|0>$ denotes the reflected state, and $|1\rangle$ the transmitted
state.  In the proposed theory, the same wholistic description of the
state is given by the base-2 normal $r(\pi/2, \lambda)$. If $r(\pi/2,
\lambda) \ge 1/2$, then under $R$, a photon will be detected if a measurement is
made in the transmitted beam, and if $r(\pi/2, \lambda) < 1/2$ then a photon
will be detected if a measurement is made in the reflected beam. (The
relation between reduction and measurement is described in section
\ref{sec:measurement} below.) However, in addition,  the real-number state
$r(\pi/2,\lambda)$ can also describe each of the component parts separately. In
this more local description, the state in the transmitted beam is given by
$r(\pi/2, \lambda)$, the state in the reflected beam is given by $r(\pi/2,
\lambda+\pi)=1-r(\pi/2,\lambda)$. Now apply the operator $R_1$ to
detect the presence of a photon (in either the transmitted or refected beam).
In this local view, if $R_1$ leads to a reduction to  $.111\ldots$ (particle
detected) in one of the beams, then the state in the other beam is not in any
way required to reduce non-locally to $.000\ldots$, but, rather, remains in its
original unreduced state. Of course, if $r(\pi, \lambda+\pi) \ge 1/2$, then
necessarily, $r(\pi/2,\lambda) < 1/2$. Hence, if a particle is detected in the
reflected beam, it can be deduced that the leading digit in the binary expansion
of $r(\pi/2,\lambda)$ is a `0', and hence a photon would not be detected were a
measurement to be made on the transmitted beam. Sampling over an ensemble of
photons defined by varying $\lambda$, then 50\%  of the photons will be detected
in the transmitted beam, and 50\% in the reflected beam. In summary, the
proposed theory is capable of giving a more local description to quantum
phenomena than standard quantum theory. The implications of this for Bell's
theorem is discussed in section \ref{sec:bell}.

Consider now the action of the second half-silvered mirror of a Mach-Zehnder
interferometer.  By time symmetry, if the global incoming state is $r(\pi/2,
\lambda)$, then the outgoing  state must be $.111\ldots$ corresponding to
particle detection in one channel only. Hence the process of
constructive and destructive interference in the two output channels,
respectively, is well described.

By contrast, suppose one of the two incoming beams in the second
half-silvered mirror has been blocked off, so that the incoming state is
$r(\pi/2, \lambda)$ in just one beam. In the proposed theory, the corresponding
outgoing global state is equal to $r$ if $r \ge 1/2$, equal to $1-r$ if $r <
1/2$. This result makes use of the property of self
similarity. Specifically, $r(\pi/2,\lambda)$ can be approximated arbitrarily
closely by some reduced number $.mmm\ldots$ in base-$2^N$ providing $N$ is
sufficiently big. However, in the proposed theory, we can treat `$m$' as just a
label (see section \ref{sec:3l}) . Hence,  if $.111\ldots$
transforms to $r(\pi,\lambda)$ under the action of the half-silvered mirror,
then the state $.mmm\ldots$ will transform to the number $r'$ whose base-$2^N$
expansion has the same form as the binary expansion of $r(\pi/2,\lambda)$, but
where the digits $m$ and $\bar{m}$ replace 1 and 0 respectively, with
$m+\bar{m}=2^N-1$. In this way, the digits in the leading $N$ places in the
binary expansion of $r'$ will be the same as in the leading $N$ places in the
binary expansion of $r$, if the binary expansion of $r$ begins with a `1' ($r\ge
1/2$). Conversely, the digits in the leading $N$ places in the binary expansion
of $r'$ will be the $\phi$ transform of the digits in the leading $N$ places in
the binary expansion of $r$, if $r < 1/2$. Either way, sampling  over an
ensemble of photons defined by varying $\lambda$, then 50\% of photons would be
detected in the transmitted beam, and 50\% in the reflected beam.

\section{The Three-Level Quantum State}
\label{sec:3l}

In this section, using the concepts and constructions discussed above, the
real-number representation of the 3-level quantum state is defined. From this,
extension to a general $M$-level system is, in principle, straightforward.

In standard quantum theory, the 3-level state vector
 \be
|\psi\rangle=\alpha_0 |0\rangle + \alpha_1 |1\rangle + \alpha_2 |2\rangle
 \ee
is an element of a complex 3-dimensional Hilbert space. Here we describe the
probabilistic equivalence  \be |\psi\rangle \lhd r(\theta_1, \theta_2, 
\lambda_1, \lambda_2) \ee where, as before, $0 \le r \le 1$ and the equivalence 
is with respect to some particular observable. The six degrees of freedom in the 
three complex numbers $\alpha_j$, less normalisation and the choice of global 
phase factor $\exp{i\gamma}$, leads to the dependence of $r$ on four arguments. 
As before the choice of $\gamma$ is equivalent to the specific choice of base-3 
normal number $r_0$ such that \be
\label{eq:r0}
\frac{1}{\sqrt{3}}(|0\rangle+|1\rangle+|2\rangle) \lhd r_0
\ee
One possible choice for $r_0$  is the (base-3) Champernowne number
$r_C=.012101112\ldots$.

\subsection{Reduced Numbers and Dynamical State Reduction}

As with the 2-level state, the reduced numbers $0 \le r_{\perp j} \le 1$, $j \in
\{0,1,2\}$ can be defined by deleting all occurrences of the digit `$j$' in the
base-3 expansion of $r$. For example, using the base-3
Champernowne number, $r_{C\ \perp 0}=.1211112\ldots$. Note that if $r$ is base-3
normal, then $r_{\perp j}$ is base-2 normal. These reduced numbers define
quantum sub-systems. For example, \begin{eqnarray}
\label{eq:base2}
\frac{1}{\sqrt{2}}(|0\rangle+|1\rangle) &\lhd& r_{0\ \perp 2} \nonumber \\
\frac{1}{\sqrt{2}}(|1\rangle+|2\rangle) &\lhd& r_{0\ \perp 0} \nonumber \\
\frac{1}{\sqrt{2}}(|2\rangle+|0\rangle) &\lhd& r_{0\ \perp 1}
\end{eqnarray}
Extending this notion by deleting multiple digits, we have three
base-1 normals \begin{eqnarray}
\label{eq:base1}
|0> \lhd r_{0\ \perp \{1,2\}}&=&(r_{0\ \perp 1})_{\perp\ 2}=.000\ldots
\nonumber \\ |1> \lhd r_{0\ \perp \{2,0\}}&=&.111\ldots \nonumber \\
|2> \lhd r_{0\ \perp \{0,1\}}&=&.222\ldots
\end{eqnarray}
Associated with these reduced numbers, we can define three irreversible
dynamical reduction operators $R_j$ for the 3-level system (similar to the
qubit). Specifically, with base-3 expansions,  \begin{eqnarray}
 \label{eq:reduction}
R_j&:& .a_1a_2a_3 \ldots \mapsto .jjj\ldots \ \mathrm{if}\
a_1=j \nonumber \\
R_j&=&\mathrm{Identity}\ \mathrm{if}\  a_1 \ne j
\end{eqnarray}
the latter being equivalent to a so-called `null' reduction in standard
quantum theory. The compound reduction operator $R=R_0R_1R_2$ allows us to ask:
under compound measurement, would the real-number state reduce to level 1, 2 or
3? Again, the reader is referred to section \ref{sec:measurement} for a
discussion on the relation between reduction and measurement.

 \subsection{The Phase Function $r(\lambda_1, \lambda_2)$}

Consider now a two-dimensional real-valued function
$r(\lambda_1, \lambda_2)$ associated with the equivalence
\be
\label{eq:3level}
\frac{1}{\sqrt{3}}(|0\rangle+e^{i\lambda_1}(|1\rangle+e^{i\lambda_2}|2\rangle) )
\lhd r(\lambda_1,\lambda_2)
\ee
which in the proposed theory is a base-3 normal number, and generalises the
corresponding base-2 normal $r(\lambda)$ for the qubit. As
discussed below, the set of reals $r(\lambda_1, \lambda_2)$ where $\lambda_1$ is
a triadic rational multiple of $\pi$ and $\lambda_2$ is a dyadic rational
multiple of $\pi$, defines the sample space on which quantum measurement
probabilities are defined.

As before, $r(\lambda_1, \lambda_2)$ is constructed from a family of permutation
operators which have complex structure. Let
\be
\mathcal{S}_r=\{a_1,a_2,a_3\ldots\}
\ee
with $a_i \in \{0,1,2\}$, define the base-3 expansion $r=.a_1a_2a_3\ldots$ of
some base-3 normal $0 \le r \le 1$. For the 3-level system, put $\phi(0)=1,\
\phi(1)=2,\ \phi(2)=0$ and define
\be
\omega_{(3)}^{1/3}(\mathcal{S}_r)=\{\phi(a_3),a_1, a_2, \phi(a_6), a_4, a_5
\ldots \} \ee
whence
\begin{eqnarray}
\omega_{(3)}(\mathcal{S}_r)&=&\{\phi(a_1), \phi(a_2), \phi(a_3) \ldots \}
\nonumber \\
\omega_{(3)}^2(\mathcal{S}_r)&=&\{\phi^2(a_1), \phi^2(a_2), \phi^2(a_3) \ldots
\} \nonumber \\
\omega_{(3)}^3(\mathcal{S}_r)&=&\{\phi^3(a_1), \phi^3(a_2), \phi^3(a_3) \ldots
\}=\mathcal{S}_r
\end{eqnarray}
so that $\omega_{(3)}$, $\omega_{(3)}^2$ and $\omega_{(3)}^3$ are
pemutation-operator representations of third roots of unity. For consistency,
the $i$ operator from the previous section can be relabelled as
\be
i=\omega_{(2)}^{1/2}
\ee
As with $\omega_{(2)}^{1/2}$, higher-order fractional powers of
$\omega_{(3)}^{1/3}$ are constructed by assuming self-similarity. Hence, for
example \be
\omega_{(3)}^{1/9}(\mathcal{S}_r)=\{\phi(a_9),a_7, a_8,\ a_1,a_2,a_3,\
a_4,a_5,a_6  \ldots \}
\ee
whence it is straightforward to show that
\be
\omega_{(3)}^{1/9}*\omega_{(3)}^{1/9}*\omega_{(3)}^{1/9}(\mathcal{S}_r)
=\omega_{(3)}^{1 / 3 } (\mathcal{ S } _ r )
\ee
and so on. In this way $\omega_{(3)}^q$ can be defined for any triadic rational
$q$. (And, more generally, $\omega_{(p)}^q$ for any $p$-adic rational $q$.)
Based on this, the real number transformations induced by $\omega_{(3)}$ are,
for example, \be
\tilde{\omega}_{(3)}^{1/9}(r)=.\phi(a_9)a_7a_8a_1a_2a_3a_4a_5a_6
\ldots  \ee
Putting these construction together, then
\be
\label{eq:phase}
\frac{1}{\sqrt{3}}(|0\rangle+e^{i\lambda_1}(|1\rangle+e^{i\lambda_2}|2\rangle) )
\lhd r(\lambda_1,\lambda_2)=
\tilde{\omega}_{(3)}^{3\lambda_1/2\pi}
 (\tilde{\omega}_{(2)}^{2\lambda_2/2\pi}(r_{0 \perp 0})\oplus .000\ldots)
 \ee
where $\lambda_1/2 \pi$ is a triadic rational, and $\lambda_2/2 \pi$
is a dyadic rational. The symbol `$\oplus$' means that `$0$' digits are
added to the base-3 expansion of $\tilde{\omega}_{(2)}^{2\lambda_2/2\pi}(r_{0
\perp 0})$ in exactly the same places that the same `0' digits were removed
from the base-3 expansion of $r_0$ by the $\perp 0$ operator. In more expanded
language, the right hand side of equation \ref{eq:phase} can be explained as
follows: take the base-3 expansion of $r_0$, and permute among the places where
the digits `1' and `2' occur with
$\omega_{(2)}^{2\lambda_2/2\pi}=i^{4\lambda_2/2\pi}$, keeping the places where
the digit `0' occurs unchanged. Then permute the three digits in all the
resulting places according to the permutation operator
$\omega_{(3)}^{3\lambda_1/2\pi}$. This function is defined on a
countable set of points (where $\lambda_1$ is a triadic rational multiple
of $\pi$ and $\lambda_2$ is a dyadic rational of $\pi$.)

A crucial property of this construction is that all  $r(\lambda_1,\lambda_2)$
are base-3 normal (providing  $r_0$ is). Hence we can define a rather trivial
(ie constant) probability on the sample space of real-number states
spanned by $(\lambda_1, \lambda_2)$. In particular, the probability that
$r(\lambda_1, \lambda_2)$ lies between 0 and 1/3, or between 1/3 and
2/3, or between 2/3 and 1, is equal to 1/3. Hence, from the reduction
equation \ref{eq:reduction}, the probability that $R_j$ takes $r(\lambda_1,
\lambda_2)$ to the reduced real-number state $.jjj\ldots$ is also equal to 1/3.

\subsection{Application of The Reduction Procedure}

Consider some fixed values of $(\lambda_1, \lambda_2)$, which without loss of
generality, may as well be $(0,0)$. The (partial) reduction procedure defined
in section \ref{sec:nonnormal} is now used to define the equivalence
 \be
 |\psi \rangle=\cos \frac{\theta_1}{2} |0\rangle +\sin
\frac{\theta_1}{2}(\cos\frac{\theta_2}{2}|1\rangle + \sin
\frac{\theta_2}{2}|2\rangle) \lhd r(\theta_1, \theta_2, 0,0).
\ee
For example, with $\theta_1=\pi$
\be
|\psi\rangle=\cos\frac{\theta_2}{2}|1\rangle +\sin
\frac{\theta_2}{2}|2\rangle \lhd r(\pi, \theta_2, 0,0)
\ee
can be defined using the reduction procedure
given
 \begin{eqnarray}
 \label{eq:project}
|1\rangle &\lhd& r(\pi,0,0,0)=.111\ldots \ \mathrm{base\ 1\ normal}\nonumber
\\ \frac{1}{\sqrt{2}}(|1\rangle+|2\rangle) &\lhd& r(\pi,
\frac{\pi}{2}, 0, 0) =.a_1a_2a_3\ldots \ \mathrm{base\ 2\
normal}\nonumber \\
|2\rangle &\lhd& r(\pi,\pi,0,0)=.222\ldots \ \mathrm{base\ 1\ normal}
 \end{eqnarray}
Note that the expansions of all the numbers in equations \ref{eq:project} 
contain the digits `1' and`2', rather than `0's and `1's. To apply the 
reduction procedure as in section \ref{sec:nonnormal}, first relabel the digits, 
so that $a_j \mapsto a_j-1$. Then, as before, starting with the (relabelled) 
sequence $\mathcal{S}=\{a_1,a_2,a_3\ldots\}$, if $a_j=1$, delete 
$a_j$ from $\mathcal{S}$ iff  \be \label{eq:pp3} .a_ja_{j+1}a_{j+2}\ldots < 
\cos^2 \frac{\theta}{2} \ee and if $a_j=0$, delete $a_j$ iff \be
\label{eq:pp4}
.a_ja_{j+1}a_{j+2}\ldots \ge \cos^2 \frac{\theta}{2}
\ee
Now return to the original digits  `1' and `2' with $a_j \mapsto a_j+1$ and 
relabel the surviving sequence of digits as $\{a'_1,a'_2,a'_3\ldots\}$. Finally,
 \be \label{eq:ddd}
 r(\pi,\theta_2,0,0)=.a'_1a'_2a'_3\ldots
 \ee
In this way, all the real-number states in
equations \ref{eq:base2} and \ref{eq:base1} are defined.

Now consider
\be
|\psi\rangle=\sqrt{\frac{1}{3}}|0\rangle+\sqrt{\frac{2}{3}}(\cos
\frac{\theta_2}{2}|1\rangle + \sin \frac{\theta_2}{2}|2\rangle \lhd
r(\theta^*, \theta_2,0,0) \ee
where $\cos \theta^*/2=1/\sqrt{3}$. Using equation \ref{eq:ddd}, this can
be defined by
 \begin{eqnarray}
 r(\theta^*, \theta_2,0,0)&=&.a'_1a'_2a'_3\ldots  \oplus .000\ldots\nonumber \\
&=&b_1b_2b_3\ldots
 \end{eqnarray}
where, as before, the $\oplus$ operator means - add the digits `0' into the
expansion $.a'_1a'_2a'_3\ldots$ in exactly the places they were removed from the
expansion of $r_0$ by the $\perp 0$ operator.

Finally, apply the reduction procedure to define $r(\theta_1,\theta_2,0,0)$
given $r(0,\theta_2,0,0)=.000\ldots$ which contains only the digits `0',
$r(\theta^*, \theta_2,0,0)=.b_1b_2b_3\ldots$ which contains all three digits,
and $r(\pi, \theta_2,0,0)=.a'_1a'_2a'_3\ldots$ which contains only the digits
`1' and `2'. As before, start with the sequence
$\mathcal{S}=\{b_1,b_2,b_3\ldots\}$. Let $c_j=1$ if $b_j=1$ or $b_j=2$, and let
$c_j=0$ if $b_j=0$. Then if $c_j=1$, delete $b_j$ from $\mathcal{S}$ iff \be
.c_jc_{j+1}c_{j+2}\ldots \le \cos^2 \frac{\theta}{2}
\ee
and if $c_j=0$, delete $b_j$ from $\mathcal{S}$ iff
\be
.c_jc_{j+1}c_{j+2}\ldots > \cos^2 \frac{\theta}{2}.
\ee
The digits that survive these deletions are relabelled as
$\{b'_1,b'_2,b'_3\ldots\}$ and
\be
 r(\theta_1,\theta_2,0,0)=.b'_1b'_2b'_3\ldots.
 \ee
This completes the description of the equivalence
\be
 |\psi \rangle=\cos \frac{\theta_1}{2} |0\rangle +\sin
\frac{\theta_1}{2}e^{i\lambda_1}(\cos\frac{\theta_2}{2}|1\rangle +
e^{i\lambda_2}\sin \frac{\theta_2}{2}|2\rangle) \lhd r(\theta_1, \theta_2,
\lambda_1, \lambda_2). \ee

\subsection{Equivalence with the Trace Rule}

As previously discussed, sampling over $\lambda_1$ and $\lambda_2$ defines, in
the proposed theory, the probability measure $P$ for quantum measurement
outcome. For example, we can ask, what is the probability that the base-3
expansion of the real-number state $r(\theta_1, \theta_2, \lambda_1, 
\lambda_2)$, for randomly-chosen $(\lambda_1, \lambda_2)$ begins with a
`0', '1'or a `2'. This can be straightforwardly calculated using the method
discussed in section \ref{sec:nonnormal}. Hence, for example, entirely
equivalent to the result derived in section \ref{sec:nonnormal}, the probability
that the base-3 expansion of the partially-reduced number $r(\pi, \theta_2,
\lambda_1, \lambda_2)$ begins with a `0' is equal to 0, that it begins with a
`1' is equal to $\cos^2 \theta_2/2$, and that it begins with a `2' is equal to
$\sin^2 \theta_2/2$.

More generally, defining the three functions $\rho_j(\theta_1,
\theta_2)$ which give the probability that  $r(\theta_1, \theta_2,
\lambda_1, \lambda_2)$ is attracted to $\mathcal{A}_j$ under $R_j$, then  from
the construction in section \ref{sec:nonnormal}  \begin{eqnarray}
\rho_0&=&\cos^2 \theta_1/2\nonumber \\
\rho_1&=&\sin^2\theta_1/2\cos^2\theta_2/2\nonumber \\
\rho_2&=&\sin^2 \theta_1/2 \sin^2 \theta_2/2
\end{eqnarray}
consistent with the trace rule for measurement outcome in standard quantum
theory.

From a number-theoretic perspective, the $\rho_j(\theta_1, \theta_2)$ define
the degree of normality of the number $r(\theta_1, \theta_2,
\lambda_1, \lambda_2)$ - specifically, $\rho_j$ is a measure of the fraction of
places in the base-3 expansion of $r$ where the digit $j$ occurs, compared with
the fraction of places where any of the other permitted digits occur. In this
way, the probability density matrix of quantum theory is subsumed, 
in the proposed theory, by the number-theoretic degree of normality of the
corresponding real-number state.

\section{Sub-Systems, Measurement Outcome and the Classical
Limit} \label{sec:measurement}

Consider an $M$-level system $\mathrm{Sys}_M$ where $M \gg 1$. Imagine, for
example, $\mathrm{Sys}_M$ represents the entire universe. In
subsection \ref{sec:reduction}, it was speculated that degree of Borelian
normality of the real-number state $r$ of an isolated system, will be
monotonically decreasing due to the effect of number-theoretic reduction
operators, representing the ubiquitous effects of quantum self-gravitation.
Hence, in the proposed theory, the real-number $\mathcal{O}$-state $r$ of the
initial big bang could be imagined to be an (optimally-normal) base-$M$ normal
real, such as  the base-$M$ Champernowne number $.012\ldots M'M101112\ldots$
where $M'=M-1$. The cosmic state would then evolve deterministically though a
sequence of base-$N$ normal reals, $N<M$, to some gravitationally-clumped
minimally-normal real-number state $r=jjj\ldots$, where $j \in \{0,1,2\ldots,
M'\}$. This time-asymmetric evolution is entirely consistent with
the Weyl curvature hypothesis, as discussed by Penrose (1989, 1994).

We define an $m$-level sub-system $\mathrm{Sys}_m \subset
\mathrm{Sys}_M$, $m < M$, by making use of the concept of reduced numbers
defined in section \ref{sec:3l}. In particular, choosing $m$ digits from the
original $M$, then the state of the $m$-level  sub-system is
merely given by the base-$M$ expansion of $r$, retaining only the chosen digits.
For example, if  \be |0\rangle+|1\rangle+|2\rangle+|3\rangle+|4\rangle+|5\rangle
+|6\rangle+|7\rangle \lhd .0532017621435\ldots
 \ee
represents the state of an 8-level system, then \be |2\rangle+|5\rangle \lhd
r_{\perp \{0,1,3,4,6,7\}}=.5225  \ldots \ee
denotes the state of a qubit subset of this system.
As a description solely of a qubit, without consideration of the larger system
to which it belongs, the digits `2' and `5' are merely arbitrary (but
ordered) symbols representing the two different levels of the qubit. Hence we
could replace $2 \mapsto 0$ and $5 \mapsto 1$, so that $.5225\ldots  \mapsto
.1001\ldots$ and use the results from sections \ref{sec:complex}-
\ref{sec:dynamics} to describe the dynamics of this qubit sub-system.

 However, the reduced  real-number state $.222\ldots$ and $.555\ldots$  also
characterise specific levels of the larger 8-level system. Hence, conversely, in
the definition of the Bloch-sphere equivalent of the real-number state space of
the qubit,  the real-numbers $r=.000\ldots$ and $r=.111\ldots$ should be
thought of as associated with some reduced real-number states $.mmm\ldots$ and 
$.m'm'm'\ldots$ of $\mathrm{Sys}_M$ of which the qubit is a sub-system. It is 
the relation of these reduced states with respect to the $M$ possible reduced 
states of $\mathrm{Sys}_m$, that determines the absolute orientation of the 
polar axes of the qubit's Bloch-sphere equivalent.

In this way, the process of measurement can be discussed without any
recourse to an arbitary classical/quantum split as is required in standard
quantum theory (Bell, 1990, 1993).  In the proposed theory, a measuring system 
is some  $M$-level system with attractors $\mathcal{A}_j$ where $r=.jjj\ldots$ 
and $j \in \{0,1,2\ldots M-1\}$. As discussed below, it will be important that 
$M \gg 1$. However, in addition, a crucial aspect of any measurement process is 
the notion of detector gain - a small signal arising from the process of state 
reduction is amplified to give a large output signal (eg in a photomultiplier). 
In the proposed model, the notion of detector gain can be described by imagining 
the design of the measuring system to be such that for $J \le j < K$, a state
inititially at $\mathcal{A}_j$ will necessarily cascade through the sequence
$\mathcal{A}_{j+1} \rightarrow \mathcal{A}_{j+2} \rightarrow \mathcal{A}_{j+3}
\ldots \rightarrow \mathcal{A}_K$.  We presume that the cascade process is
sufficiently extensive that the real-number states $r=.JJJ\ldots$ and 
$r=.KKK\ldots$ are perceptibly different from one another to a human observer. 
For $j < J$, imagine the real-number state $r=.jjj\ldots$ as stationary.

With $j'=j-1$, we presume that $M$ is sufficiently large that the real-number state
$r=.j'j'j'\ldots$ and any real-number state whose
base-M expansion comprises only the digits $j$ and $j'$, are
not perceptibly different from one another.  Imagine  such a
measuring system  initially in the (reduced) stable real-number state 
$r=.J'J'J'\ldots$, where $J'=J-1$.  The system is then perturbed by a qubit 
(say) in the unreduced real-number state $.a_1a_2a_3\ldots$ where $a_j \in 
\{0,1\}$. In the case of optimal detector efficiency, the evolution of the 
measuring system will be tightly coupled with that of the qubit - in this 
situation we can call the two real-number states entangled. Hence, by the 
arguments above,  the real-number state $r=.b_1b_2b_3 \ldots$ where $b_j=J'$ if 
$a_j=0$, $b_j=J$ if $a_j=1$ will represent both the real-number state of the 
qubit (under the symbol replacement $J' \mapsto 0,\ J \mapsto 1$), and the 
real-number state of the measuring system to which the qubit is tightly coupled. 
By definition, the real-number state $r=.b_1b_2b_3\ldots$ of the perturbed 
measuring system is not perceptibly different from its initial real-number state 
$r=.J'J'J'\ldots$. If $a_1=1$, then, under reduction, the real-number state of 
the measuring system evolves to $r=.JJJ\ldots$ which then, by the 
detector gain process discussed above, cascades to $r=.KKK\ldots$. 
Alternatively, if $a_1=0$, the real-number state of the measuring system is 
unchanged.

This notion of entanglement between the single qubit and the larger measuring
system, implies that the time for the qubit to attract
to one of the fixed points $.000\ldots$ or $.111\ldots$ would correspond to the
time it takes the perturbed state $r=.b_1b_2b_3\ldots$  of
the measuring system to be attracted to either $.J'J'J'\ldots$ or $.JJJ\ldots$.
It would seem plausible that this timescale would be proportional to
the density of reduced real-number states $.jjj\ldots$ in state space. Following  
ideas developed in Di\'{o}si (1989) and Penrose (1994, 1998),
the density of reduced states might be related in some way to the difference in
gravitational self-energy asssociated with such pairs of reduced states.
Motivated by the geometric ideas in the theory of general relativity, suppose
the curvature of the Bloch-sphere equivalent was determined by the density of
reduced states of the system $\mathrm{Sys}_M$ to which the qubit was
entangled, and that $\alpha$ in equation \ref{eq:ode} depended on the radius of
curvature of the Bloch-sphere equivalent. Then the greater the density
of reduced states associated with $\mathrm{Sys}_M$,  the shorter the timescale
for attraction of the qubit state to one of the fixed points. If the
appropriate measure of interaction between the qubit and $\mathrm{Sys}_M$ is
gravitational, then, in the proposed theory, the `basis' problem for the qubit
is solved in much the same way that Mach's principle solves the  `inertial
frame' problem in general relativity. Clearly such ideas require further
development.

In the more general case where the evolution of the qubit and
the measuring system $\mathrm{Sys}_M$ is not so tightly coupled, then the digits
$b_i$ and $a_i$ would not be in simple one-to-one correspondence. In this case,  
the fraction of digits where $b_j=J'$ would be larger than the fraction of 
digits where $b_j=J$, by some factor depending on the degree of entanglement. As 
a result,  there will be  a greater likelihood that the corresponding 
measurement will be null. We discuss further this notion of weak measurement in 
section \ref{sec:stochastic} below.

In this discussion it has not been necessary to make any arbitrary
classical/quantum split: the process of measurement is, in principle, contained
within the theory. However, it is possible to refer to a `classical limit'
for the proposed theory: one where evolutionary laws are derived solely on the
basis of transformations between the sub-set of reduced real-number state $.iii\ldots$.
This limit clearly makes sense when the set of such reduced real-number state is
comparatively dense in state space, ie when $M$ is large.  Suppose $r(t)$
represents the exact evolution of $r$. Then, in the proposed theory, the
classical limit corresponds to finding the best approximation $r_c(t) \sim
.jjj\ldots$ where $j=j(t)$.  For example, the cascade $\mathcal{A}_{j+1}
\rightarrow \mathcal{A}_{j+2} \rightarrow \mathcal{A}_{j+3} \ldots \rightarrow
\mathcal{A}_K$ as described above is a classical process (as it would be in 
conventional physics). By labelling the reduced state $.jjj\ldots$ by the 
integer $j$, then, in the proposed theory,  the laws of classical physics would 
correspond to evolutionary  equations defined by relationships between the 
integers (or, equivalently, the ordinary rationals of computational physics, ie 
without concern to normality).

\section{Composite Systems: The  Quantum Computer and The Navier-Stokes
Computer} \label{sec:computing}

In sections \ref{sec:3l} and \ref{sec:measurement} we discussed the use of
reduced numbers to determine the real-number states of sub-systems. The rule 
for defining the real number state of a composite system is the simple converse. 
For example, if we have the real-number binary expansion equivalent
 \be
\frac{(|0> + e^{i \lambda_n} |1\rangle)}{\sqrt{2}} \lhd
r(\pi/2, \lambda_n) = .a^{(n)}_1a^{(n)}_2a^{(n)}_3 \ldots
\ee
for any one qubit, then, in the proposed theory,  the real-number equivalent
$r^{(N)}$ of the $2^N$-level $N$-qubit composite
\be \label{eq:qft}
|\psi_N\rangle = \frac{(|0> + e^{i \lambda_1} |1\rangle)
(|0> + e^{i \lambda_2} |1\rangle)(|0> + e^{i \lambda_3}|1\rangle)\ldots (|0> +
 e^{i \lambda_N} |1\rangle)}{2^{N/2}}
\ee
will be the real number with the base-$2^N$ expansion
 \be
r^{(N)}=.b_1b_2b_3\ldots
\ee
where $b_j \in \{0,1,2\ldots 2^N-1\}$ and the binary expansion of $b_j$ is
\be
b_j=a^{(1)}_ja^{(2)}_ja^{(3)}_j \ldots a^{(N)}_j.
\ee
$N$ qubit Hilbert-space states, such as in equation \ref{eq:qft} are
used, for example,  in the quantum Fourier transform (eg Nielsen and Chuang,
2000), an essential part of many quantum computational algorithms.
The best classical algorithm for computing discrete Fourier transforms on
 $2^N$ elements  uses $O(N2^N)$ gates, whilst the quantum Fourier transform
 requires $O(N^2)$ gates. One of the gates in this count is
 \be
U_N=\left( \begin{array}{cc}
1& 0 \\
0& e^{2\pi i / 2^N} \\
\end{array}
\right)
\ee
acting on one of the qubit channels. In the proposed theory, the equivalent
$\tilde{U}_N$ gate corresponds to an application of the operator $i^{4/2^N}$
acting on the bit string associated with the binary expansion of the real-number
state of this qubit. On  a digital computer,  the estimation of $\tilde{U}_N$,
as discussed in section \ref{sec:complex}, would require individual operations
on the each of the first $2^{N-1}$ places in the binary expansion of $r$.  It is
this ability to perform by a single phase gate, a process that requires
exponentially many permutations on a  classical digital computer, that describes
the power of the quantum  computer in the proposed theory.

Is it conceivable, in a single-universe world view, that a deterministic 
real-number system can execute, what would correspond to $O(2^N)$  
primitive operations on a digital computer, in fixed finite time (independent of 
$N$)? In fact, there is a well-known classical system which does precisely this: 
a system based on the Navier-Stokes quations for a three dimensional turbulent 
fluid in the limit of large Reynolds number. (Indeed studies of the properties 
of these equations led the author to formulate the proposed theory!)

In the inviscid limit, scaling arguments for homogeneous isotropic turbulence
(in the Kolmogorov inertial range) suggest that information contained in
some arbitrarily small-scale `eddy' can propagate up-scale to affect the
evolution of some large-scale `eddy' of interest in finite time (essentially the
large-scale eddy turn-over time $T_1$; Lorenz, 1969; Palmer, 2000). Specific
non-differentiable solutions of the Navier-Stokes equations with this property
have  been found (Shnirelman, 1997), though rigorous generic theorems are still
lacking. More quantitatively, imagine decomposing a three-dimensional fluid
(eg streamfunction) field onto some real (Galerkin) basis, so the field is
represented by a string $\{x_1, x_2, x_3\ldots x_N\}$ of real numbers. The
number $x_j$ is the projection coefficient of the field onto the $j$th basis
function, which describes the $j$th `eddy'. By Kolmogorov scaling, the time
it takes for the $N$th eddy to affect the first (largest) eddy,  scales as
$T_1(1-\exp(-N))$. On a digitial computer, however, the time taken to perform a
computation of the evolution of $x_1$ over this same timescale, will grow as
$N^4$ (allowing for the three dimensionality of the eddies, and the need to
reduce the computational time step as the minimum eddy size decreases). Hence, a
Navier-Stokes computer (ie a turbulent fluid!), performs this calculation
exponentially more efficiently than does a digital computer.

Of course, like a quantum computer, such a Navier-Stokes
computer can only perform certain computations (based on solutions of the
Navier-Stokes equations!). However, in this respect, it is worth noting that,
through the nonlinear Hopf-Cole transformation (eg Whitham, 1974), the Burgers'
form of the Euler equation can be cast in the form of a Schr\"{o}dinger
equation, though with a real, rather than complex, state vector. It is also
interesting to note that the inviscid Navier-Stokes equations (the Euler
equations) have an interesting geometric interpretation: they describe geodesics
on the group of diffeomorphisms of the fluid 3-space.

The author's permutation construction in section 2 captures this sense of the
upscale transfer of fluid-dynamical information, and, as shown, is certainly
non-differentiable. Hence, in terms of the proposed theory,  the exponential
speed up of the quantum computer is conceptually no different from
the exponential speed up of the Navier-Stokes computer, over a conventional
digital computer. On this basis, it is suggested that the power of the quantum
computer can be explained without recourse to multiple universes, or indeed
superposed Hilbert-space eigenstates (cf Deutsch, 1997).

Just as viscosity limits the maximum size $N$ of a Galerkin expansion
of the fluid state, and hence of the maximum  processing power of a
Navier-Stokes computer, so one can question whether there  are fundamental
physical constraints which might limit the maximum possible length $2^N$ of the
binary expansion of $r$, and hence the processing power of a quantum computer.
Possibly self-gravitation of the composite $N$ qubit may provide such a
constraint, eg leading to a minimum angular resolution of the points on the
Bloch-sphere equivalent over which the real-number states are defined (see 
section \ref{sec:complex}).

\section{Bell's Theorem, Beables and Counterfactual Indefiniteness}
\label{sec:bell}

Consider a source of  pairs of entangled qubits measured by devices with
relative orientation $\Delta \theta \ne 0$. Define the coordinate system
$(\theta, \lambda)$ and poles $(p_N, p_S)$ associated with the real-number 
state space of the left-hand qubit,  and coordinate system $(\theta', \lambda')$ 
with poles $(p_{N'}, p_{S'})$ for the right-hand qubit. The co-latitude of 
$p_{N'}$ with respect to $(\theta, \lambda)$ is $\Delta \theta$. As demonstrated 
in subsection \ref{sec:epr}, the proposed theory can readily account for the 
observed correlation statistics associated with entangled EPR states. In 
subsection \ref{sec:local}, the reason why the proposed model violates Bell's 
inequalities is discussed.

\subsection{EPR-state Correlations}
\label{sec:epr}

Imagine  an ensemble $i=1,2,3\ldots N$ of pairs of qubits such that the 
real-number state of the  $i$th left-hand qubit is $r^{(i)}$, and the 
real-number state of the $i$th right-hand qubit is $r'^{(i)}$. In the proposed 
theory, the fact that these qubits are entangled implies a deterministic 
relationship between $r^{(i)}$ and $r'^{(i)}$.

To define this relationship, first write the binary expansion of $\cos^2(\Delta
\theta/2)$ as \be \cos^2(\Delta \theta/2)=.d_1d_2d_3\ldots
\ee
and the binary expansions of $r^{(i)}$ and $r'^{(i)}$ as
\begin{eqnarray}
r^{(i)}&=&.a^{(i)}_1a^{(i)}_2a^{(i)}_3\ldots \nonumber \\
r'^{(i)} &=& .a'^{(i)}_1a'^{(i)}_2a'^{(i)}_3\ldots
\end{eqnarray}
Second, decompose the set $I=\{1,2,3,\ldots N\}$ of natural numbers  into
the disjoint subsets \be
I=\bigcup I_j
\ee
where, for $j \le N$,
\be
I_j=\{i: i=2^{j-1}+(k-1)2^j;\ k \le N, i \le N\}
\ee
For example, for N=12,
\be
I=\{1,3,5,7,9,11\} \cup \{2,6,10\} \cup \{4, 12\}
\ee
For large $N$, the subset $I_j$ contains a fraction $1/2^j$ of the $N$ integers.

The required relationship between $r^{(i)}$ and $r'^{(i)}$ is defined
as follows. For $i \in I_j$, let
\be \label{eq:ent}
a'^{(i)}_l=(1-d_j) a^{(i)}_l + d_j \phi(a^{(i)}_l)
\ee
for $l=1,2,3 \ldots$. That is to say, for $i \in I_j$, and hence for a
fraction $1/2^j$ of the $N$ ensemble pairs,  $a'^{(i)}_l=\phi(a^{(i)}_l)$ if
$d_j=1$, and $a'^{(i)}_l= a^{(i)}_l$ if  $d_j=0$.
Taken over all the disjoint subsets $I_j$, the total
fraction of occasions  when the $a'^{(i)}_l= \phi(a^{(i)}_l)$ is
\be
d_1/2+d_2/2^2+d_3/2^3 \ldots
\ee
which, by definition, is equal to $\cos^2(\Delta \theta/2)$. The corresponding
 fraction of occasions where the $a'^{(i)}_l = a^{(i)}_l$ is therefore equal to
 $\sin^2(\Delta \theta/2)$.
Now if $a'^{(i)}_1=\phi(a^{(i)}_1)$,  then if $r^{(i)}$ reduces to $p_N$,
$r'^{(i)}$  reduces to $p_{S'}$, and vice versa. If we write $x_i=1$ when
$a'^{(i)}_1=a^{(i)}_1$, $x_i=-1$ otherwise,  then it is  immediate that
\be
\frac{1}{N}\sum_{i=1}^{N} x_i = \frac{1}{2}(\sin^2 \frac{\Delta \theta}{2}-
\cos^2 \frac{\Delta \theta}{2})= -\cos \Delta \theta
\ee
as required by experiment and standard quantum theory.

\subsection{The Bell Inequalities}
\label{sec:local}

Since the proposed theory is able to account for quantum correlations between
entangled EPR states, statistics from the proposed theory
violate the Bell inequalities.  Manifestly, this implies that the proposed
theory cannot be formulated as a local non-contextual theory. What is the key 
property of the proposed model that allows it to violate Bell's theorem? Let us 
use Bell's (1964) notation to describe a local non-contextual hidden variable 
form $A(\hat{\mathbf{n}}, \lambda)=\pm 1$ and $B(\hat{\mathbf{n}}, \lambda)= \pm 
1$, for predicting, respectively, the outcomes of measurement of (for example) 
spin along direction $\hat{\mathbf{n}}$ for the left and right hand 
(spin-1/2) qubits of some entangled pair. Bell's theorem requires that \be 
\label{eq:reality} B(\hat{\mathbf{n}}, \lambda)=-A(\hat{\mathbf{n}}, \lambda) 
\ee In words, equation \ref{eq:reality} requires that if the left-hand qubit was 
measured as spin up (with respect to $\hat{\mathbf{n}} \equiv p_{N}p_{S}$), then
the right-hand qubit  would have been measured as spin down, had it been
measured with respect to $p_{N}p_{S}$. In the proposed theory, equation 
\ref{eq:reality} would certainly hold if $p_Np_S=p_{N'}p_{S'}$, by the 
property of real-number states corresponding to antipodal points. However, this 
hypothetical measurement on the right-hand qubit is counterfactual, since, by 
assumption,  the right-hand qubit was actually measured with respect to $p_{N'}\ 
p_{S'} \ne p_{N}p_{S}$.

Consider a pair of entangled qubits, as described in subsection
\ref{sec:epr}. Let the  real-number state of the left-hand qubit be given by
$r(p)$, the real-number state of the the right hand qubit be given by $r'(p')$.
Then the counterfactual measurement on the right-hand qubit, as defined in the
previous paragraph, has definite outcome only if $r(p')$
is well defined. With probability one it is not. To see this, note that
according to the relationship established in subsection \ref{sec:epr}, $p$ and
$p'$ are not (when $p_{N'}\ p_{S'} \ne p_{N}p_{S}$) antipodal points. Now since
$r'(p')$ is well defined, $p'$ must (from section \ref{sec:complex}) lie on a
meridian (call it $\Lambda'_j$), belonging to the set $\Lambda'$ of meridians
whose longitudes are dyadic rational multiples of $\pi$.  From section
\ref{sec:nonnormal}, $r'(p')$ is defined on the continuum of all possible points
on $\Lambda'_j$. However, by the discussion at the end of section
\ref{sec:complex}, $\Lambda'_j$ is intersected only a countable number of times
by the set of meridians $\Lambda$, emanating from $p_{N}$, on which the
real-number state $r(p)$ is well defined. Hence the probability that $p'$ lies
on a meridian in $\Lambda$, is equal to zero. Hence with probability one, the
counterfactual real-number state $r(p')$ is undefined - it is therefore
\emph{not} a beable. (Equivalently, $r'(p)$ is also undefined with probability
one). The fact that equation \ref{eq:reality} is neither true nor false in the
proposed theory is reminiscent of non-Boolean logic in certain topos theories
(Isham, 1997).

Counterfactual indefiniteness is not a property of conventional classical
deterministic (eg chaotic) dynamics. For example, using a numerical weather
prediction model (Palmer, 2000), it  is straightforward to estimate
(in the model) what the weather in London would  have been like today if the
temperature in Chicago had been two degrees colder  a week earlier. In the
proposed theory, this would not be possible; if a Turing  machine was programmed
to estimate $r'(p)$ recursively, given $r(p)$, $r'(p')$, and the
constructions of the proposed theory, then, with probability one,  it would
never halt. In this sense, one could view $r'(p)$ as an uncomputable element of
the theory. This notion that quantum physics may have uncomputable elements has
been discussed by Penrose (1989, 1994) - the real-number states $r'(p)$ in the
proposed theory are uncomputable, because, \emph{a fortiori}, they cannot be
defined from the state space of the two entangled qubits.

These arguments apply to microscopic qubits, to macroscopic $M$-level
measuring apparati and indeed to experimenters themselves. Hence, it is
similarly meaningless to ask what would have been the real-number
state of the measuring apparatus had the experimenter chosen a
different measurement orientation to the one actually chosen (again, the
proposed theory would declare the real-number state to be undefined).
Indeed, the same is true of the state of the experimenter herself; effectively,
the proposed theory compromises the intuitive notion of experimenter `free-will'
(cf Bell, 1985), at least at some fundamental level, though this does not imply
that the proposed theory conflicts with the experimenter's perception that her
choices of measurement orientation are freely chosen.

The lack of well-defined counterfactual-measurement outcomes is one of the
primary reasons why the real-number states in the proposed theory are beables.
On the one hand, for any mathematically well-defined real-number state,
we can estimate how it will transform  under reduction (and therefore
measurement). On the other hand, counterfactual measurements which by
definition cannot be elements of reality, are not associated with well-defined
real-number states. Hence real-number states and elements of reality are in one-to-one
correspondence in the proposed theory.

As mentioned, the proposed theory cannot be both local and non-contextual. We
have argued that counterfactual indefiniteness implies that non-contextuality is
violated. Indeed, the proposed theory is local, at least in the original EPR
sense. We have argued in section \ref{sec:unitary} when discussing the
individual measurements on the two channels of an interferometer, that the
process of performing a measurement which reduces the real-number state
in one channel, is not required to induce a change in the real-number state of
the other channel. The same argument applies when discussing measurements on
entangled EPR real-number states - a measurement on the left-hand qubit need
not imply a change in the real-number state of the right-hand qubit. The
proposed theory is therefore consistent with the notion of relativistic invariance.

 \section{Weak Reduction and Stochastic Quantum Theory}
\label{sec:stochastic}
 
Consider the reduction of some qubit to either $p_N$ or $p_S$ to be
broken up into a large number of `weak reductions'
\be
p \mapsto p_1 \mapsto p_2 \mapsto p_3 \cdots,
\ee
where each step, $p_j \mapsto p_{j+1}$,
corresponds to a partial reduction towards either some  $p_{N_j}$ or the
antipodal $p_{S_j}$. Each weak reduction will be defined by a partial
integration of equation \ref{eq:ode}. Specifically, let  $\theta_j$, $\lambda_j$
denote the co-latitude and longitude of $p_j$ with respect to a coordinate
system with poles $p_{N_j}$, $p_{S_j}$. Let $r(p_j)$ denote the qubit real-number state with
respect to $(\theta_j, \lambda_j)$. Then $p_j \mapsto
p_{j+1}$ can be expressed by the finite difference equations
\ref{eq:ode}  \begin{eqnarray}
 \label{eq:stoch1}
 \theta_{j+1}&=&\theta_j + \alpha (r-\frac{1}{2})\sin \theta \Delta t\nonumber
\\
\lambda_{j+1}&=&\lambda_j
  \end{eqnarray}
Before the next step is implemented, there must be a transformation  to the
coordinate system based on $p_{N_{j+1}}\ p_{S_{j+1}}$, giving \begin{eqnarray}
\label{eq:stoch2}
\lambda_{j+1}&\mapsto &\lambda_{j+1} + \delta \lambda_{j+1}\nonumber \\
\theta_{j+1}&\mapsto &\theta_{j+1} + \delta \theta_{j+1}\nonumber
\end{eqnarray}
As discussed in section \ref{sec:complex}, the evolution $r(\lambda)
\mapsto r(\lambda+\delta \lambda)$, for small $\delta\lambda$ would be
interpreted in conventional computational analysis as white-noise stochastic. 
The sequence of weak reductions associated with the $\theta$ evolution in 
equation \ref{eq:stoch1} therefore appears as if generated by a random 
(gambler's ruin) walk (Pearle, 1993) towards the attractors $p_{N}$, $p_{S}$. In 
view of the relationship between reduction and measurement outcome, one could 
think of this sequence of reductions as being associated with measurement by a 
detector whose orientation was not precisely fixed.

Hence, this deterministic model of weak reduction appears to be consistent,
for all practical purposes, with stochastic  quantum theory (Percival, 1998).
However, in stochastic quantum theory, an additional source of stochastic noise
is prescribed,  and the Schr\"{o}dinger equation is  generalised to take the
form of a stochastic differential equation in Hilbert  space. However, in the
proposed theory,  there are no stochastic sources. Hence at a fundamental level, 
it is claimed that the proposed deterministic theory is intrinsically simpler 
than stochastic quantum theory.

\section{Discussion}
\label{sec:discussion}

A theory has been developed, where the quantum state is not, axiomatically, an
element of a Hilbert space over the complex numbers (and hence not in a
complex linear superposition of eigenstates), but is a single real number $0 \le
r \le 1$.  A key number-theoretic notion that provides substance to this (at
first sight preposterous) notion is that of number-theoretic Borelian normality;
in the proposed theory, the real-number state corresponding to
maximally-superposed $M$-level Hilbert-space state is (a good rational
approximation to) a base-$M$ normal real, whilst eigenstates of the
Hilbert-space (for a particular observable) correspond to the non-normal
(trivially, base-1 normal) reals $.000\ldots$, $.111\ldots$, $.222\ldots$ and so
on to $.M'M'M'\ldots$, where $M'=M-1$. One of the key features of the proposed
theory, which extends quantum theory, is that the sample space over which
quantum measurement probabilities can be calculated, is precisely defined.
Specifically, this space is defined by the set of base-$M$ normal numbers
generated by a family of self-similar permutation operators acting on the digits
and places in the base-$M$ expansion of $r$. These permutation
operators have complex structure (being a representation of the $M$th roots of
unity) and subsume the essential role of complex numbers in standard quantum
theory. In the proposed theory, real-number state reduction is a precisely
deterministic process based on the application of number-theoretic
operators which reduce the degree of normality of $r$. It is speculated that
these number-theoretic reduction operators describe the (irreversible) process
of gravitation at the quantum level. Through the use of these reduction
operators, the relation between real-number state reduction and measurement
outcome can be defined, without recourse to an arbitrary classical/quantum
split. From the degree of number-theoretic normality of $r$, one can
directly infer the trace rule for measurement outcome in standard quantum
theory. Both observers and the observed are defined from the elements of the
theory.

As mentioned, the proposed deterministic realistic theory is more than a
reformulation of standard quantum theory, it is an extension of quantum
theory, specifically in relation to the measurement problem. It has been shown
that many of the foundational difficulties of standard quantum theory are
much less problematic in the proposed theory. However, this doesn't mean that
the proposed theory is more like classical theory than standard quantum
theory. One of the most profound differences between the proposed theory and
classical deterministic theory concerns the notion of counterfactual
indefinitess. This (emergent) property is shown to have profound implications
for the interpretation of Bell's theorem, and allows the proposed theory to be
local in the sense of EPR. As such, unlike standard quantum theory,
the model is entirely consistent with relativistic invariance.

Since measurement outcomes certainly correspond to elements of reality, all
real-number states correspond (through the reduction operators) to elements of
reality. Conversely all counterfactual measurement outcomes, which by definition
cannot correspond to elements of reality, have undefined real-number states.
Moreoever, within the proposed theory, observers, measuring apparati (and indeed
the whole cosmos) also have real-number states albeit with very large degrees
of normality (eg corresponding to base-$M$ normal reals, where $M \gg 1$.
Putting these facts together, it can be seen that Bell's (1993) notion of
`beable' describes this real-number state precisely.

It can be noted that it is straightforward to show that the proposed
theory satisfies Hardy's 5 `reasonable' probability axioms for a quantum theory
- consistent with the claim that at the level of probabilities the proposed
realistic theory is equivalent to standard quantum theory. Hardy's 5th axiom,
which distinguishes quantum from classical theory, and requires probability
densities to vary continuously in state space, is satisfied by virtue of the
fact that the degree of normality varies continuously in real-number state
space.

This work was motivated by the author's studies of  predictability in
meteorology, especially of three-dimensional high Reynolds-number turbulence in
the inertial sub-range (Palmer, 2000). The existence of a finite-time
predictability horizon in such turbulent systems is fundamentally different
from the infinite-predictability horizon associated with conventional chaotic
systems. This paradigm has been used in section \ref{sec:computing} to define
the concept of a `Navier-Stokes' computer, to illustrate that the exponential
speed up of certain quantum computations need not imply a many-worlds
interpretation.

\section*{Acknowledgement}
My thanks to Lucien Hardy for a helpful discussion on the
formulation of the proposed theory with respect to the 3-level system, and to
Michael McIntyre for encouragement and many helpful comments on how to
improve the paper's lucidity.

 \section*{References}

\begin{description}

\item Bell, J. S., 1964: On the Einstein-Podolsky-Rosen paradox.
Physics 1, 195-200.

\item Bell, J.S., 1985: Free variables and local causality. Dialectica, 39,
85-96.

 \item Bell, J.S., 1990: Against `Measurement'. Physics World, 3, 33-40.

\item Bell, J.S., 1993: Speakable and unspeakable in quantum mechanics.
Cambridge University Press. 212pp.

 \item Deutsch, D., 1997: The Fabric of Reality. Penguin Books. 390pp.

\item Di\'{o}si, L., 1989: Models for universal reduction of macroscopic quantum
fluctuations. Phys.Rev. A, 40, 1165-1174.

 \item Duane, G.S., 2001: Violations of
Bell's inequality in synchronized  hyperchaos. Foundations of Physics Letters,
14, 341-353.

\item Einstein, A., Podolsky, P. and N.Rosen, 1935: Can quantum-mechanical
 description of physical reality be considered complete? Phys.Rev., 47, 777-780.

\item Hardy, G.H. and Wright, E.M., 1979: The Theory of Numbers.
Oxford University Press.

\item Hardy, L., 2001: Quantum theory from five reasonable axioms. 
quant-ph/0101012.

\item Isham C., 1997:Topos Theory and Consistent Histories: The Internal Logic
of the Set of all Consistent Sets. Int.J.Theor.Phys., 36, 785-814

\item Kent, A. 2002: Locality and reality revisited. quant-ph/0202064.

\item Lorenz, E.N., 1969: The predictability of a flow which
possesses many scales of motion. Tellus, 21, 289-307.

\item Nicolis, J.S., Nicolis, G. and Nicolis, C., 2001: Nonlinear
dynamics and the two-slit delayed experiment. Chaos, Solitons and
Fractals, 12, 407-416.

\item Nielsen, M.A. and I.L.Chuang, 2000: Quantum Computing and Quantum
 Information. Cambridge University Press. 676pp

\item Ott, E., Sommerer, J.C., Alexander, J.C., Kan, I., and
J.A.Yorke, 1993: Scaling behavior of chaotic systems with riddled
basins. Phys. Rev. Lett., 71, 4134-4137.

\item Palmer, T.N., 1995: A local deterministic model of quantum
spin measurement. Proc.R.Soc.Lond. A, 451, 585-608.

\item Palmer, T.N., 2000: Predicting uncertainty in forecasts of weather and
 climate. Reports on Progress in Physics., 63, 71-116.

 \item Pearle, P., 1993: Ways to describe dynamical state vector reduction. 
Phys. Rev., A48, 913-923.
  
 \item Penrose, R., 1989: The Emperor's New Mind. Oxford 
University Press.  Oxford. 466pp

\item Penrose, R., 1994: Shadows of the mind. Oxford University
Press. Oxford. 457p

\item Penrose, R., 1998: Quantum computation, entanglement and state reduction. 
Phil. Trans. Roy. Soc., A451, 1927-1939.
 
 \item Pitowsky, I., 1983: Deterministic 
model of spin and
statistics. Phys.Rev., D27, 2316-2326.

\item Prigogine, I. and Y. Elskens, 1987: Irreversibility, stochasticity and
 non-locality in classical dynamics. In `Quantum Implications: Essays in Honour
 of David Bohm'. Routledge. London. 455pp

\item Shnirelman, A., 1997: On the nonuniqueness of weak
solutions to the Euler equation. Comm. Pure \& Appl. Math., 50,
1260-1286.

\item 't Hooft, G., 1999: Quantum gravity as a dissipative deterministic system.
quant-qc/9903084

\item Whitham, G.B., 1974: Linear and Non-linear waves. John
Wiley. New York. 636pp.

\end{description}

\end{document}